\documentclass[useAMS,usenatbib,usegraphicx]{mn2e}

\usepackage{journals}
\usepackage{url}
\usepackage{multirow}
\newcommand{\thickhline}{\noalign{\hrule height 0.8pt}}
\usepackage{threeparttable}
\title[Optical and NIR constraints on 6 LMXBs]{Sample of LMXBs in the Galactic bulge. I. Optical and near-infrared constraints from the Virtual Observatory}
\author[Zolotukhin, and Revnivtsev]{Ivan Yu. Zolotukhin$^{1,2}$\thanks{E-mail:
iz@sai.msu.ru (IZ)}, and Mikhail G.
Revnivtsev$^{3,4}$\\
$^{1}$Sternberg Astronomical Institute, Moscow State University, Universitetskij pr., 13, 119992, Moscow, Russia\\
$^{2}$Observatoire de Paris-Meudon, LERMA, UMR~8112, 61 Av. de l'Observatoire, 75014 Paris, France\\
$^{3}$Excellence Cluster Universe, Technische Universit\"at M\"unchen, Boltzmannstr.2, 85748 Garching, Germany\\
$^{4}$Space Research Institute, Russian Academy of Sciences, Profsoyuznaya 84/32, 117997 Moscow, Russia\\
}
\begin{document}

\date{Accepted 2010 Sep XX. Received 2010 Jun XX; in original form 2010 May XX}

\pagerange{\pageref{firstpage}--\pageref{lastpage}} \pubyear{2010}

\maketitle

\label{firstpage}

\begin{abstract}
We report on the archival optical and near-infrared observations of 6 low mass X-ray binaries situated in the Galactic bulge. We processed several recent {\it Chandra} and {\it XMM-Newton} as well as {\it Einstein} datasets of a binary systems suspected to be ultracompact, which gave us arcsec-scale positional uncertainty estimates. We then undertook comprehensive search in existing archives and other Virtual Observatory resources in order to discover unpublished optical/NIR data on these objects. We found and analysed data from ESO Archive and UKIRT Infrared Deep Sky Survey (UKIDSS) on SLX~1735-269, 3A~1742-294, SLX~1744-299, SLX~1744-300, GX~3+1, IGR~J17505-2644 systems and publish their finding charts and optical flux constraints in this paper, as well as simple estimates of the physical parameters of these objects.
\end{abstract}

\begin{keywords}
X-rays: binaries -- infrared: stars -- stars: individual: SLX 1735-269 -- stars: individual: 3A 1742-294 -- stars: individual: SLX 1744-299 -- stars: individual: SLX 1744-300
\end{keywords}

\section{Introduction}

Low mass X-ray binaries (LMXBs) are the most common astrophysical objects, that host neutron stars and black holes. Galactic LMXBs were discovered at the dawn of X-ray astronomy \citep{giacconi62} and now with the help of advanced orbital X-ray facilities like {\it Chandra} or {\it XMM-Newton} they are being intensively studied in other galaxies as well \citep{fabbiano89, primini93, sarazin03, kim04, gilfanov04, fabbiano07}.

Apart from being hosts for exotic objects like black holes and neutron stars, LMXBs attract a lot of attention as compact binaries. Indeed, understanding of their secular evolution can give us insights about the rate of events extremely important in astrophysics, like: 1) SN Ia, standard cosmology candles; 2) mergers of compact objects (like white dwarf-white dwarf, neutron star-neutron star), which are crucial for our understanding of gravitational wave signals and construction of future gravitational wave detectors \citep{belczynski02, nelemans09}.

Orbital periods of LMXBs evolve very slowly, making it challenging to observe their change\citep{burderi09}. However, it is clear that secular evolution of long living LMXBs directly influences overall statistical properties of their population, like distribution over orbital periods or X-ray luminosities \citep{rappaport83, kolb93, howell01}. Therefore, by measuring statistical properties of galactic LMXBs, one can make important conclusions about mechanisms of their long-term evolution. The ultimate sample for this purpose is the set of {\it persistent} sources, because for these sources we can make estimate of their time averaged mass transfer rate with more or less confidence, which is almost absent in the case of transient objects. However, even in the case of persistent sources there might be also uncertainties connected with sources' long time-scale variability, see e.g. \citet{priedhorsky87, gilfanov05, maccarone10}.

In order to link the properties of binary systems with their statistical distributions, we need to measure main parameters of LMXBs, such as orbital periods, type of donor star, and others. These detailed studies are only possible for systems within our Galaxy. But even for them it has not yet been completed in a systematic manner, while considerable efforts were invested in such projects (see e.g., \citet{bandyopadhyay99, jonker00, wachter05b}). At the moment the chance to know LMXB orbital parameters strongly increases if the binary system harbours giant companion or if it is seen edge-on (and thus we observe periodic dips or eclipses of X-ray source).

To continue our previous efforts of X-ray binaries identification \citep{zolotukhin_atel_09, zolotukhin10a}, we undertake systematic study of unbiased sample of {\em persistent} LMXBs in optical and near infrared spectral range. The sample of persistent LMXBs is taken from the deep survey of the Galaxy, performed by {\it INTEGRAL} observatory in energy range 17-60~keV \citep{krivonos07,revnivtsev08}, which is free from all complications, caused by the interstellar absorption in the Galactic disc. As the optical and NIR emission of the LMXBs is mainly caused by the reprocessing of their X-ray flux in the extended accretion disk (see e.g., \citealt{paradijs94}; however, at longer wavelengths there might be a considerable contribution from the jet, see e.g. \citet{homan05, russel07, migliari10} for comparison of the NIR brightness of sources with their X-ray luminosity), it will allow us to make at least an estimate of the size of the accretion disk, and therefore the binary system orbital parameters. In this work we study sources in the direction of Galactic Centre, most of them very likely reside in the Galactic bulge (see e.g. discussion in \citealt{revnivtsev08}). This gives an additional advantage over some arbitrarily chosen set of LMXBs to adopt a canonical distance of 8~kpc in case of missing distance estimate for a source.

As the first step of this study we performed dedicated search of the unpublished optical and NIR data publicly available for the sources of our interest in observational archives and other resources of the Virtual Observatory (VO). In order to facilitate possible follow-up studies, we publish immediate observational results for a subset of them, leaving other sources with deep enough data existing in the VO and thorough theoretical interpretation for the future.

\section{Data}

\subsection{X-ray}

The sample of LMXBs, which we study, is taken from the {\it INTEGRAL} all sky survey \citep{krivonos07}. We have limited ourselves with sources in the Galactic bulge region, which are known to be neither AGNs nor HMXBs \citep{revnivtsev08}. Accurate astrometric positions of a subsample of these sources were extracted either from archival {\it Chandra} observations, or from dedicated {\it Chandra} observations, performed in 2009 (PI M.~Revnivtsev). In all but one analysed {\it Chandra} observations, the astrometric accuracy of the sources position is provided by direct imaging. Position of sources were determined with {\sc wavdetect} task of {\sc CIAO 4.2} package and has 90~per~cent confidence level. The position of the source GX 3+1 is measured from observation, taken in Continuous-Clocking mode of {\it Chandra} (observation ID 2745), therefore essentially only one coordinate of the source position is measured with high accuracy. Position of this source was additionally constrained by archival {\it Einstein}/HRI imaging observation.

\subsection{Optical and NIR}

We have checked 2 well-known data collections, ESO Archive\footnote{\url{http://archive.eso.org}} and WFCAM Science Archive\footnote{\url{http://surveys.roe.ac.uk/wsa/}}, for the optical/NIR imaging data on the sources of our interest. Following the principal differences between these data sources, we employed slightly different processing sequences.

In spite of the difficulties with the calibration data that frequently accompany proper ESO Archive datasets reduction, we accurately retrieved them and reduced raw data (both science and Landolt standards frames) to the science-ready form using generic routines from {\sc IRAF ccdproc} package. Then we performed PSF photometry measurements of science frames and aperture photometry of the standards fields using {\sc IRAF daophot} package, thus transforming instrumental magnitudes to the standard (Vega) system.

WFCAM Science Archive was used to retrieve publicly available science-ready $JHK$ imaging data of the Galactic Plane Survey (GPS), which is a part of UKIRT InfraRed Deep Sky Survey (UKIDSS) \citep{lawrence07}, Data Release 3. All frames were binned $2 \times 2$ getting 0.4~arcsec~pix$^{-1}$ scale to get rid of evident CCD artefacts. In order to ensure the search for NIR counterparts with the maximum sensitivity we decided to employ the same photometry procedures for all UKIDSS fields, as we did for ESO data. Calibration dependencies in this case were constructed comparing our 2~arcsec aperture measurements with UKIDSS $JHK$ magnitudes measured in this aperture ({\it *AperMag3} in terms of UKIDSS data columns), thus transforming our instrumental magnitudes to the intrinsic UKIDSS magnitude system. i.e. to the Vega system as for the ESO data.

In cases where we mention upper limits, they were estimated using magnitudes of the faintest stars detected at $3\sigma$ level above the background by our algorithm.

For ESO data that frequently come without astrometric calibration, we performed our own calibration procedure using {\sc SCAMP} routine \citep{bertin06}. For UKIDSS data we used in-place WCS solution considering it to be precise at 100~mas level for each coordinate with respect to 2MASS astrometry system \citep[see e.g.][]{deacon09}, which itself is known to be accurate at 100~mas level in the ICRS reference frame \citep{skrutskie06}. Our astrometric uncertainty thus includes not only our coordinates error estimate, but also uncertainty of the reference system we used in each specific case. For the reasons of interoperability with X-ray observations, by astrometric uncertainty we usually mean radius of a positional error circle unless otherwise specified.

\section{Observations and results}

\subsection{SLX 1735-269}

SLX~1735-269 was previously studied by \citet{wilson03} who obtained with {\it Chandra} its X-ray position at arcsec accuracy level and first attempted to detect its NIR counterpart using dedicated observations at UKIRT. There was no source found in the $J$ band down to the limiting magnitude of 19.4.

\begin{figure}
\includegraphics[width=0.4\textwidth]{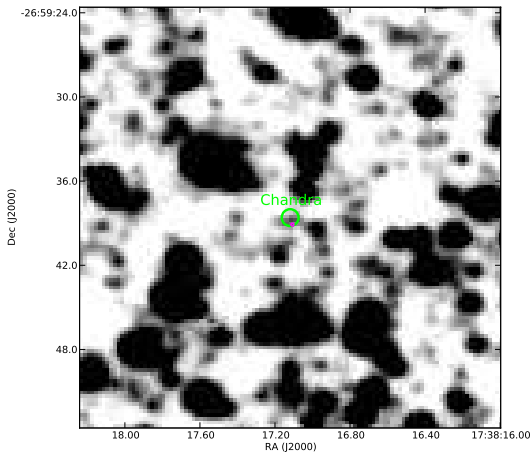}
\caption{$R$ filter image of the SLX~1735-269 field taken with 900~sec exposure using EMMI detector at ESO 3.6-m NTT telescope. Green circle denotes {\it Chandra} positional uncertainty as published in \citet{wilson03}, magenta circle marks proposed counterpart position (see the text).}
 \label{SLX1735-269_field}
\end{figure}

In this work we analysed $R$ and $I$ filter data on SLX~1735-269 available in ESO Archive under programme ID 67.D-0116(A). The dataset contained 900~sec exposures for every filter taken on May 28, 2001 between 08:22 UT and 08:58 UT with EMMI detector attached to ESO 3.6-m NTT telescope. The single source is clearly visible inside {\it Chandra} error box by \citet{wilson03} in $R$ image (see Fig.~\ref{SLX1735-269_field}) and is at $3\sigma$ level in $I$. We estimated the brightness of this positional candidate as $R = 21.31 \pm 0.12$; $I = 20.10 \pm 0.07$. The astrometric solution was obtained in the 2MASS reference frame and had 0.15~arcsec calibration uncertainty (see Table~\ref{summary_table}). Note, however, that due to the shape of the measured source we cannot rule out its double nature and this may have influenced our coordinates and magnitude estimate.

X-ray bursting behaviour of SLX~1735-269 was extensively analysed in \citet{molkov05} who confirmed previous estimates of its distance as 8.5 kpc. This, together with extinction for this line of sight taken from \citet{marshall06} ($A_K \simeq 0.5$~mag) and normal extinction law as in \citet{cardelli89}, gives us upper limits (due to uncertain nature of position-based identification) on absolute magnitudes in $R$ and $I$ for this system: $M_R \gtrsim 3.4$~mag and $M_I \gtrsim 3.4$~mag.

\subsection{3A 1742-294}

3A~1742-294 (sometimes also referred as 1A~1742-294 or 2E~1742.9‐2929) is a persistent X-ray binary which exhibits type-I X-ray bursts. The first optical finding chart of this source was presented in \citet{jernigan78}, though X-ray positional uncertainty at that time did not allow to make unambiguous conclusions. \citet{wijnands06} determined with {\it Chandra} its X-ray position accurate to arcsecond level, but very little efforts were invested in a study of this error box in optical/NIR domain.

We analysed the field around 3A~1742-294 using UKIDSS GPS DR3 images obtained by 3.8-m UKIRT telescope on Jul 18, 2006 at 09:30 with 10~sec integration time (filter $J$), at 09:38 with 10~sec integration time (filter $H$), and at 09:45 with 5~sec integration time (filter $K$). We do not detect any source well within X-ray coordinates uncertainty by \citet{wijnands06}, and therefore we publish only upper limit on NIR flux of 3A~1742-294 (see Table~\ref{summary_table}). However, we note that to the South-East immediately close to the formal error circle from {\it Chandra} there is a blended object with 2 components, either of them may turn out to be a counterpart. Our PSF photometry procedure cannot resolve them, whereas single detection lies outside of the X-ray error region.

\citet{galloway08} estimated the distance to 3A~1742-294 from its photometric radius expansion X-ray bursts as $5.8 \dots 7.5$~kpc. Adopting $d = 6$~kpc and corresponding $A_K \simeq 2.0$~mag from \citet{marshall06}, above upper limit can be translated to 3A~1742-294 absolute magnitude constraint: $M_K \gtrsim 2.0$~mag.

\begin{figure}
\includegraphics[width=0.4\textwidth]{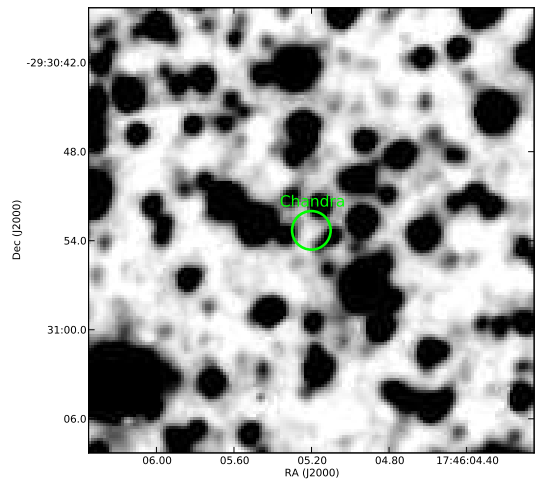}
\caption{$K$ filter image of the 3A~1742-294 field from UKIDSS GPS DR3. No sources are visible inside {\it Chandra} error box taken from \citet{wijnands06}.}
 \label{2A1742-294_field}
\end{figure}

\subsection{SLX 1744-299}

The source SLX~1744-299 is the southern source of the close pair SLX~1744-299/300 ($\sim2.5$ arcmin angular separation), discovered in the Galactic Centre region by {\it Spartan1} experiment \citep{kawai88} and resolved into two sources by {\it Spacelab} X-ray telescope \citep{skinner87, skinner90}, {\it MIR-KVANT-TTM}, and {\it GRANAT}/ART-P telescopes \citep{sunyaev91}. Both sources are known to be X-ray bursters \citep{skinner90, sunyaev91}.

Position of the source was determined with the help of {\it Chandra} imaging, performed during observations IDs 2834 (ACIS-I imaging, Oct 23, 2002) and 9106 (HRC-I imaging, Feb 7, 2008). The source, being very bright for ACIS-I capacity, suffers from strong pile-up, which creates a hole around the source position. We have used interception of this area with the readout streak, clearly visible on the image as the estimate of the centroid of the source position: RA=17:47:25.88, Dec=-30:00:02.0 (J2000). Absence of large number of X-ray sources, available for cross-matching with existing optical or IR images, does not allow us to improve the accuracy of the astrometry, therefore we anticipate that the uncertainty of the position here is mainly determined by the {\it Chandra} aspect solution and equals to $\sim$0.6~arcsec.
Second {\it Chandra} observation of the source was performed with HRC-I and does not suffer from pile-up. We have determined the position of the source with the help of {\sc wavdetect} task of {\sc CIAO 4.0} package. As the source in this observation was relatively far from the optical axis of the telescope ($\sim$3~arcmin), accuracy of its localisation is worse then usual. According to study of \citet{alexander03}, the accuracy of {\it Chandra} localisation of sources at these offsets is about $\sim0.4$ arcsec if the references of X-ray and optical images match. In the absence of cross-match between our images and the optical reference frame, we added 0.6 arcsec quadratically to the resulted uncertainty radius, thus obtaining accuracy $\sim 0.7$ arcsec.

\begin{figure*}
\includegraphics[width=0.4\textwidth]{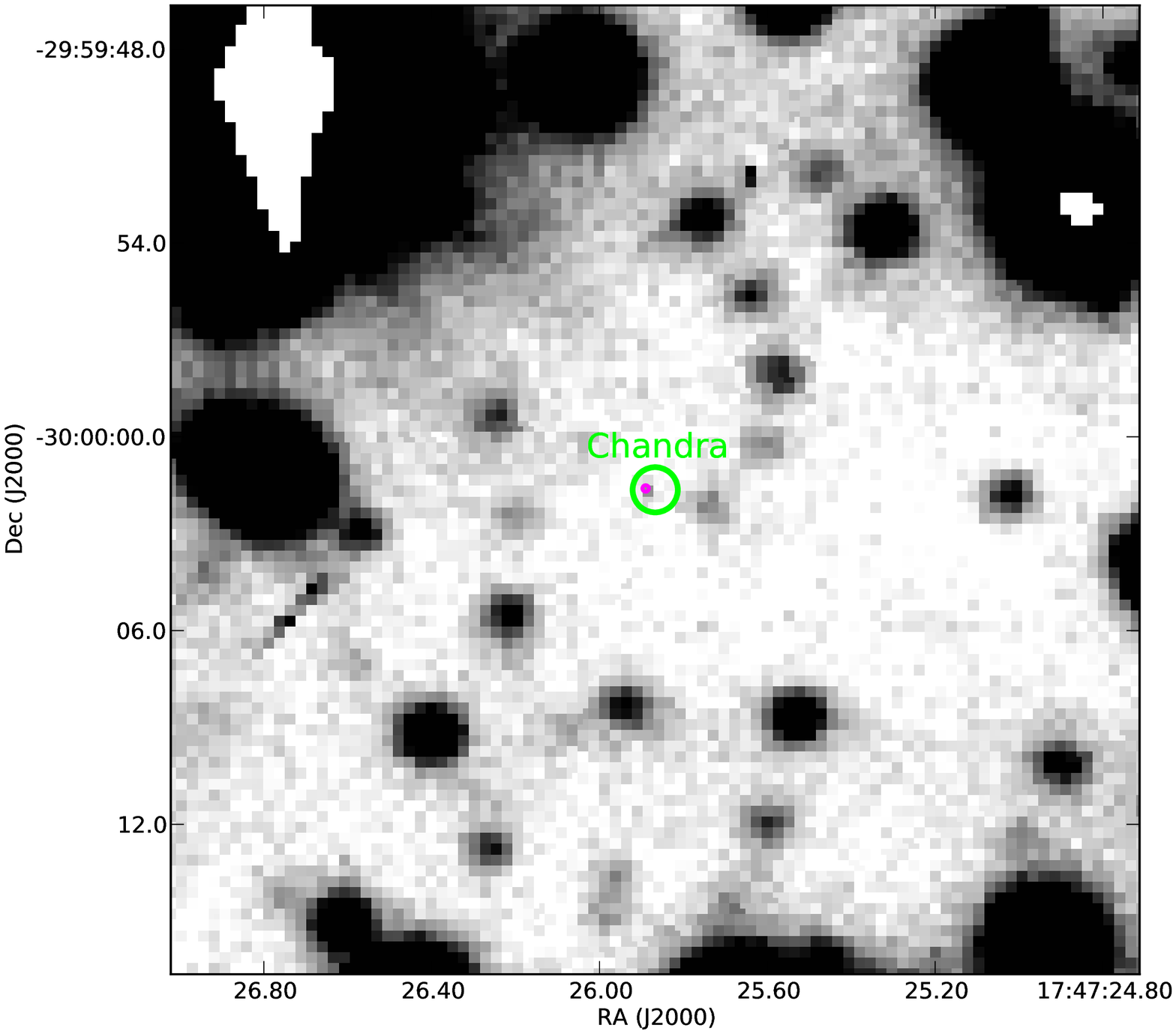}
\includegraphics[width=0.4\textwidth]{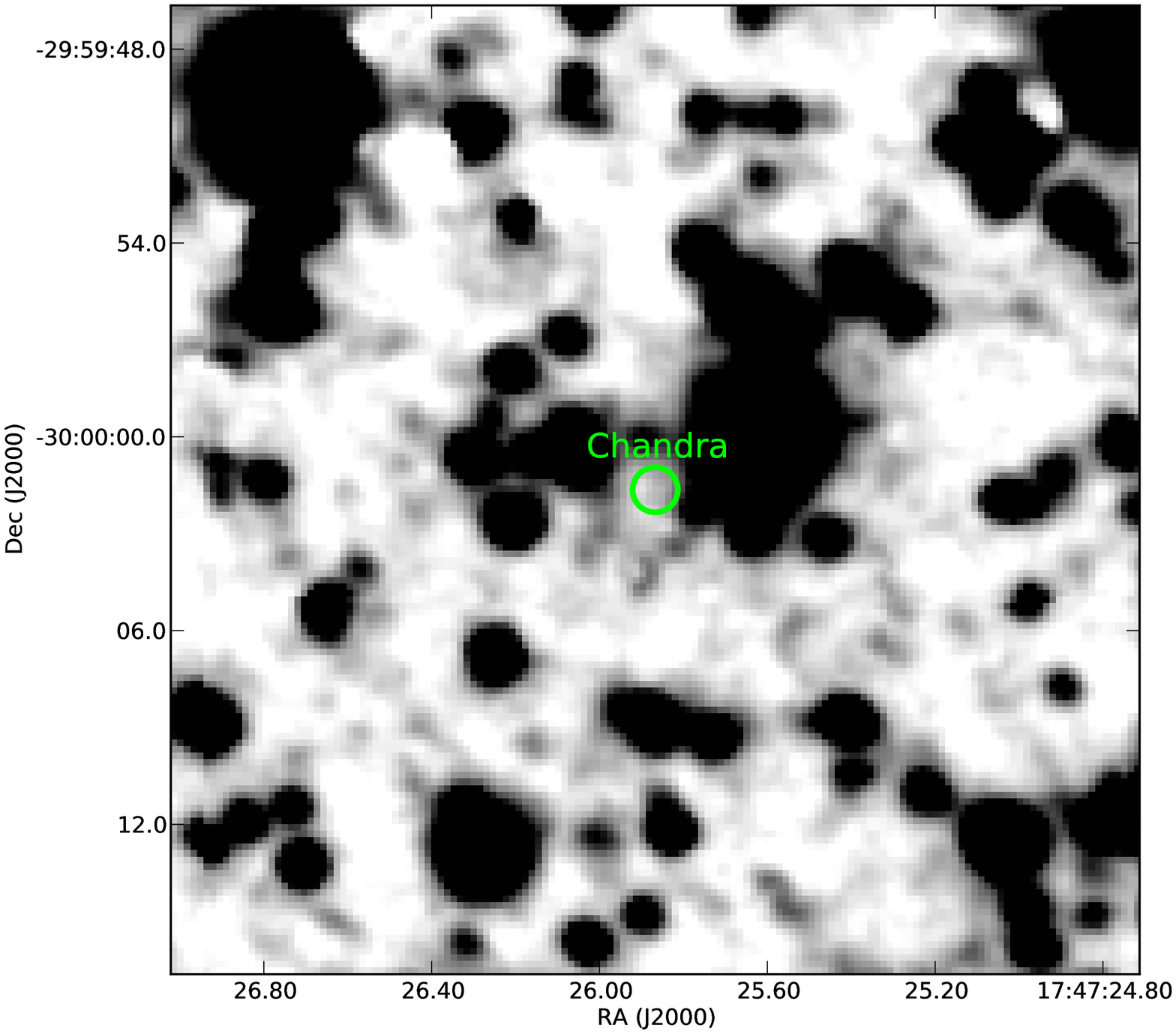}
\caption{SLX~1744-299 field with {\it Chandra} (green) positional uncertainty overplotted. {\it Left panel}: The co-added $I$ image constructed from three 600~sec exposures taken with EMMI at 3.6-m NTT telescope. Magenta circle marks proposed counterpart position. {\it Right panel}: $K$ image from UKIDSS GPS DR3. No sources are visible within X-ray error region.}
 \label{SLX1744-299_field}
\end{figure*}

We analysed $I$ filter data on SLX~1744-299 available in ESO Archive under programme ID 079.D-0385(C). The dataset contained six 600~sec exposures taken on Jun 23, 2007 between 06:27 and 07:36 UT with EMMI detector attached to ESO 3.6-m NTT telescope. First three exposures were centred at SLX~1744-299, whereas remaining ones -- at SLX~1744-300, though X-ray positions for both sources are visible on all frames as there is only 2.7~arcmin between them, and they fall well within EMMI single field of view. Observing conditions became slightly worse to the second half of the dataset, so we summed only first three of them, getting 1800~sec of the total integration time. The single source is clearly visible inside {\it Chandra} error box in the co-added image (see Fig.~\ref{SLX1744-299_field}, left panel), though we note the source to be sharp, somewhat untypically for this image. At the same time it is marginally detected at individual exposures, so we cannot immediately consider it as an artefact, especially after careful inspection of all the calibration frames we used, since they contain nothing that can influence the source's profile in an observed way. PSF photometry of these data yields $I = 23.37 \pm 0.28$ for this positional candidate. The astrometric solution was achieved to 0.2~arcsec precision in GSC 2.2 reference frame, which is known to be precise up to 170~mas level on both coordinates \citep[see e.g.][]{fienga04}.

This field is also present in UKIDSS GPS DR3 survey (see Fig.~\ref{SLX1744-299_field}, right panel). Frames taken by 3.8-m UKIRT telescope on Jul 20, 2006 between 09:16 and 09:30 UT (10~sec exposure for $J$ and $H$, and 5~sec for $K$) shows nothing remarkable in the error box, closely packed between neighbouring objects in the dense field, so we consider this as an upper limit estimate (see Table~\ref{summary_table}).

At the canonical distance of 8~kpc using the same extinction treatment as above we estimate $M_I \gtrsim 2.2$ for SLX~1744-299.

\subsection{SLX 1744-300}

The source is the northern source of the pair of X-ray bursters SLX~1744-299/300.

Position of the source was determined with the help of {\it Chandra} observation ID 9106 (HRC-I imaging). During this observation the source was quite far from the optical axis of the telescope ($\sim$5.4~arcmin), where its point spread function, and thus the accuracy of the localisation, is significantly worse. We have determined the position of the source with the help of {\sc wavdetect} task of {\sc CIAO 4.0} package. The centroid of the source position was determined to be: RA=17:47:26.02, Dec=-30:02:41.8 (J2000) with the estimated accuracy 0.7 arcsec.

\begin{figure*}
\includegraphics[width=0.4\textwidth]{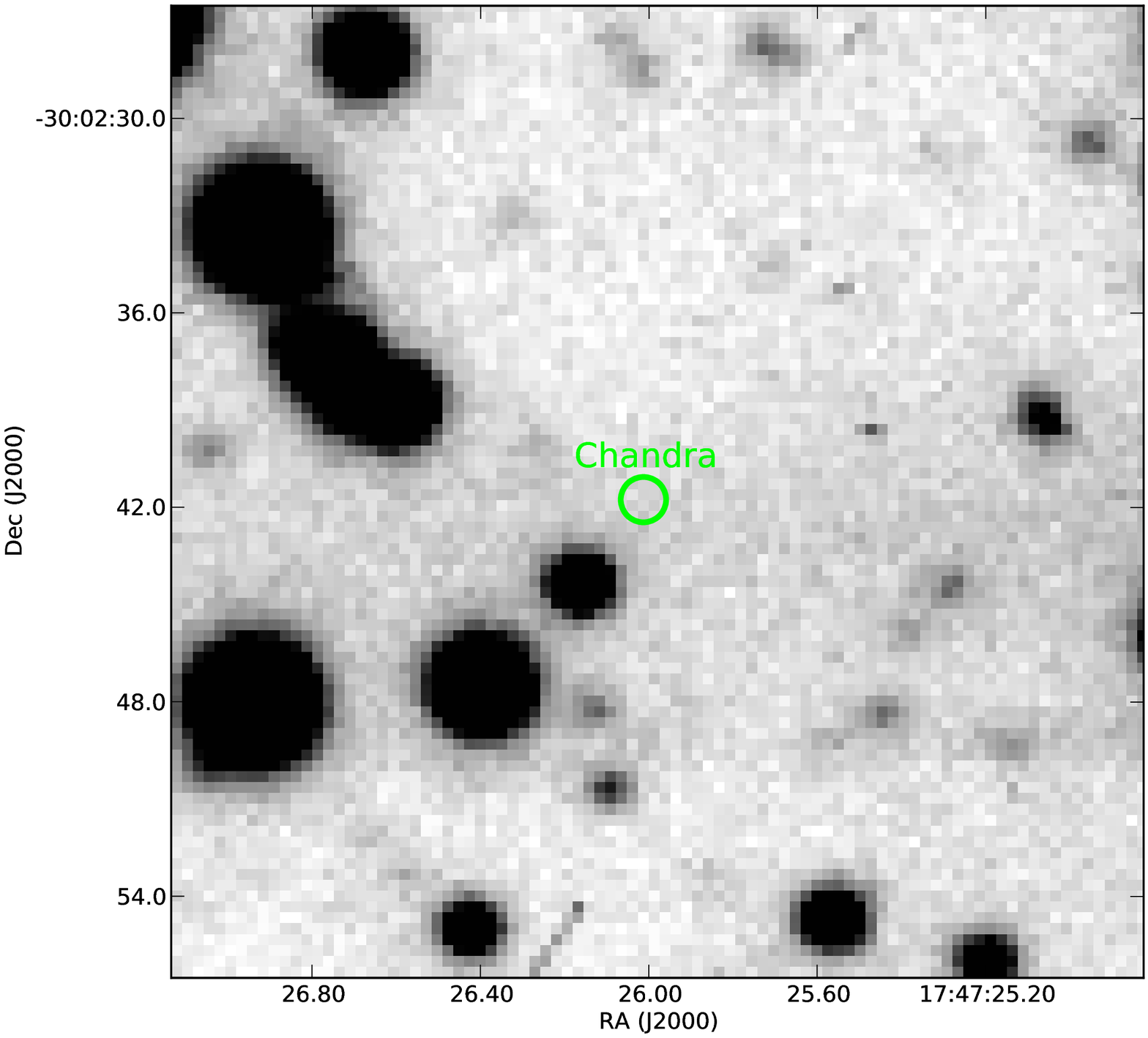}
\includegraphics[width=0.4\textwidth]{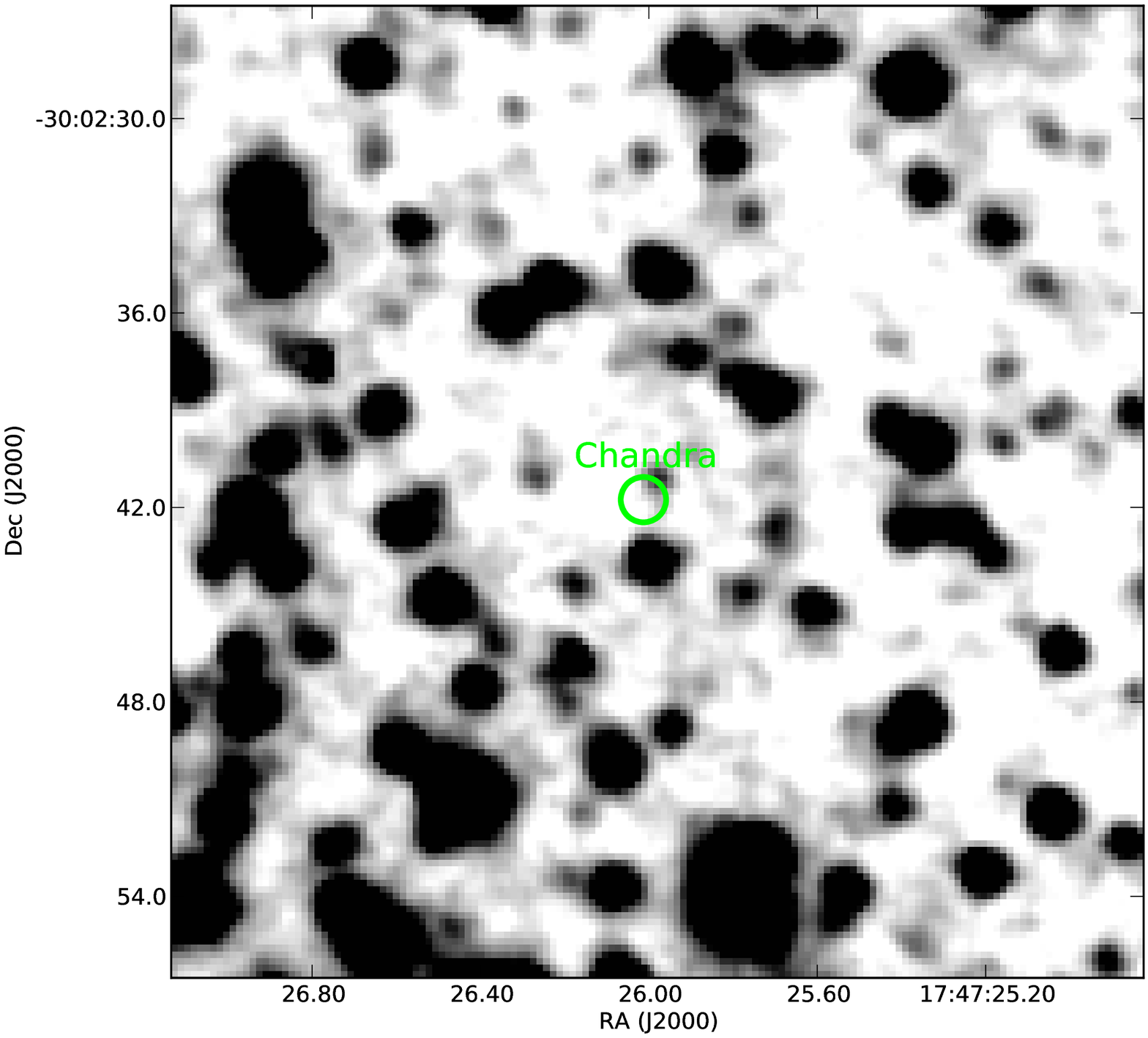}
\caption{SLX~1744-300 field with {\it Chandra} positional uncertainty overplotted. {\it Left panel}: The co-added $I$ image constructed from three 600~sec exposures taken with EMMI at 3.6-m NTT telescope; there are no sources detected within X-ray coordinates error. {\it Right panel}: $K$ filter image from UKIDSS GPS DR3.}
\label{SLX1744-300_field}
\end{figure*}

For SLX~1744-300 we used exactly the same datasets in both ESO and UKIDSS archives, as for SLX~1744-299. We do not detect any source up to $3\sigma$ limiting magnitude of $I = 23.4$ (see Fig.~\ref{SLX1744-300_field}, left panel) in ESO data. UKIDSS dataset also contains nothing remarkable in the error box (though an object RA=17:47:25.97, Dec=-30:02:41.1 (J2000) is detected 0.3~arcsec away of its boundary in $H$ and $K$ images: $H = 17.90 \pm 0.14; K = 16.75 \pm 0.14$) and we hence publish only upper limits in Table~\ref{summary_table}.

At the canonical distance of 8~kpc using the same extinction treatment as above we are able to pose the following constraint on SLX~1744-300 absolute magnitude: $M_K \gtrsim 1.9$.

\subsection{GX 3+1}

GX~3+1 is a low mass X-ray binary which is known to emit X-ray bursts, ranging from normal ones (see e.g. \citet{chenevez06}, and references therein) to the superburst \citep{kuulkers02}. Among many bursts detected, only one showed the evidence of photospheric radius expansion and hence was used to estimate the source distance as $d \sim 4.5$~kpc \citep{kuulkers00} for a hydrogen-rich neutron star atmosphere. However, analysis of thermonuclear bursts from GX~3+1 by \citet{hartog03} suggests He-rich fuel and hence somewhat larger value of $d = 6.1 \pm 0.1$~kpc, which we adopt below. First GX~3+1 NIR counterpart search was undertaken by \citet{naylor91}. They reported single source within {\it Einstein} and GX~3+1 lunar occultation error boxes intersection (see below).

\begin{figure}
\includegraphics[width=0.4\textwidth]{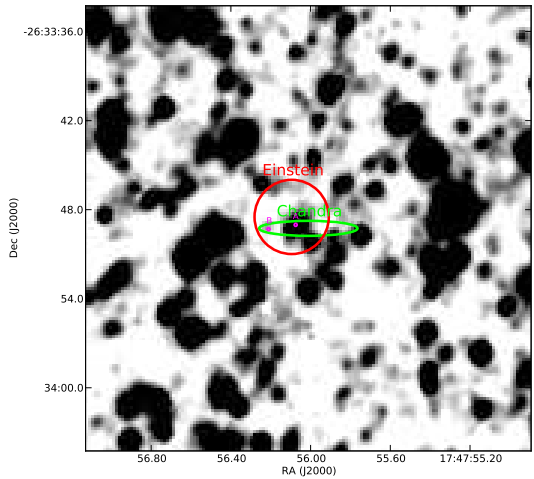}
\caption{$K$ filter image of the GX~3+1 field from UKIDSS GPS DR3 with {\it Chandra} and {\it Einstein} positional uncertainties overplotted (in green and red, correspondingly). Two possible counterparts, sources A and B, are marked by magenta circles.}
 \label{GX3+1_field}
\end{figure}

We extracted astrometric position of the source from observations of {\it Chandra} (observation ID 2745, Apr 9, 2002) and {\it Einstein}/HRI (observations Mar 23, 1979 and Mar 28, 1980). {\it Chandra} observation was performed in Continuous-Clocking mode, which precludes determination of accurate source position along one coordinate (long size of the ellipse of {\it Chandra} localisation in Fig.~\ref{GX3+1_field} is arbitrary, because position was determined only along another axis). However, combining results from {\it Einstein}/HRI (best fit position RA=17:47:56.11, Dec=-26:33:48.8 (J2000), localisation accuracy $\sim$2~arcsec) and {\it Chandra} datasets we can slightly improve the accuracy of the localisation to the level of $0.5 \times 2$ arcsec box (see Fig.\ref{GX3+1_field}).

We examined this combined {\it Chandra} and {\it Einstein} error box in UKIDSS GPS DR3 (see Fig.~\ref{GX3+1_field}). Due to its relatively large area, we detect 2 sources falling within its limits.  Its impossible to choose among them at this stage: either more compact X-ray error box is needed, or phase-resolved NIR photometry. Source A in this work most likely corresponds to the source 311 from \citet{naylor91}. The results on GX~3+1 presented in Table~\ref{summary_table} were inferred from exposures taken on May 03, 2007 between 13:47 and 14:02 UT with usual for survey integration times.

To estimate GX~3+1 absolute magnitude we took the brightest of the two positional candidates thus making our estimate an upper limit of the absolute brightness at the distance determined by \citet{hartog03}: $M_K \gtrsim 0.2$.

\subsection{IGR J17505-2644}

The source IGR~J17505-2644 is one of the faintest persistent sources in the Galactic bulge region, discovered by INTEGRAL \citep{krivonos07}. It was tentatively classified as an LMXB by \citet{revnivtsev08}.

The source was observed on Aug 5, 2009 by {\it Chandra} as a part of special programme, dedicated to measurements of astrometric positions of LMXBs (and LMXB candidates) in the Galactic bulge. Single relatively bright source was detected within source localisation circle, obtained by {\it INTEGRAL}. Position of X-ray source was measured by {\it Chandra}: RA=17:50:39.49, Dec=-26:44:36.1 (J2000), with 0.6~arcsec accuracy mainly determined by the {\it Chandra} aspect solution precision, rather than statistical accuracy of the measurements.

\begin{figure}
\includegraphics[width=0.4\textwidth]{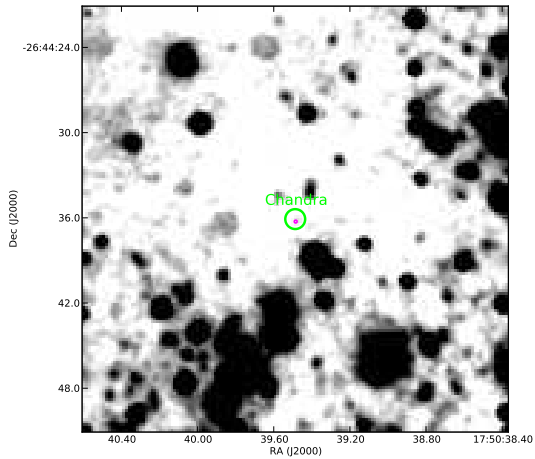}
\caption{$K$ filter image of the IGR~J17505-2644 field from UKIDSS GPS DR3 with {\it Chandra} positional uncertainty overplotted. Magenta circle denotes single source marginally visible inside X-ray error box.}
 \label{IGRJ17505-2644_field}
\end{figure}

We studied the X-ray error circle of IGR~J17505-2644 in UKIDSS GPS DR3 (see Fig.~\ref{IGRJ17505-2644_field}) in the data obtained under better than average observing conditions (with seeing $\sim$0.7~arcsec) on May 03, 2007 between 13:49 and 14:03 UT. We were not able to detect any source within X-ray coordinates uncertainty in $J$ and $H$ filters, but we marginally detect a faint source in $K$ band. We did not succeed in fitting its profile with a point-spread function due its low brightness, so we publish here its aperture photometry measurement and the astrometric position determined by simple centroids. Upper limits in $J$ and $H$ bands are significantly better then those from usual UKIDSS datasets because of the particularly good weather conditions.

Detection of a possible counterpart in $K$ band allows us to make an estimate of IGR~J17505-2644 absolute magnitude as $M_K \gtrsim 2.2$.

\section{Summary}

We have made a systematic search in ESO and WFCAM science archives for photometric images of the areas around positions of 6 LMXBs in the Galactic bulge region. In all but 2 cases we have detected faint optical/NIR sources with brightness in the range $15-23$~mag within arcsecond-scale X-ray error boxes, and obtained upper limits on optical/NIR brightness for the rest. We note that positional-based identification is always uncertain and one should take our detections with caution; they, however, still carry important information about upper limits on an optical brightness of a real counterparts. Our main results are summarized in Table~\ref{summary_table}. To quantify local star density in the vicinity of objects, we calculate simple probability of accidental background star coincidence with an error box: $P = S_{error} \times \rho$, where $S_{error}$ is an error box area and $\rho$ is an average surface density of the sources we detected in 2~arcmin field around given position. Hence, as total probability of chance star coincidence accross all error boxes is 1.64 and we detected 5 possible counterparts of 4 objects, we conclude that two or three of those objects may have true counterparts.

Assuming intrinsic colours of LMXBs with non-evolved companions as $(V-K)_0 \sim 0$ (see e.g. \citet{hertz84}) and somewhat arbitrarily, but conservative $(V-R)_0 \sim 0$ and $(V-I)_0 \sim 0$, we estimated upper limits on the orbital periods of the systems under study, presented in the last column of Table~\ref{summary_table}. We used simple scaling relations, proposed for optical emission, dominated by illumination with their central X-ray emission \citep{paradijs94}, adopting Eddington luminosity from \citet{kuulkers03}. Under these assumptions, we find that (at least) one of the studied binary systems is quite compact, with an implied orbital period less than 3 hours. We plan to make more accurate estimates of their orbital parameters in our future works.

\begin{table*}
\centering
\renewcommand{\tabcolsep}{1mm}
%\begin{minipage}{140mm}
\caption{Summary of the optical/NIR constraints, positional estimates and orbital period constraints of the sources analysed in this study. Magnitudes (either positional candidates or upper limits) are given in the Vega system. Positional errors determined for possible counterparts in this work are radii of coordinate uncertainties in arcseconds with respect to International Celestial Reference Frame. Sources of positional uncertainties are indicated in the footnotes. Probabilities of an accidental background star superposition are given for the reddest filter in case of upper limit, or the reddest filter, where detection is available (see text). Absolute magnitudes are calculated for the reddest available filter using extinction from \citet{marshall06} and distance estimates from the literature (see text); we consider all them being upper limits due to uncertain nature of position-based identification. Upper limits on orbital periods are estimated in hours using \citet{paradijs94} relation and X-ray luminosities taken from the references, indicated in the footnotes.}
\label{summary_table}
%\tiny
\begin{threeparttable}
\begin{tabular}{cccccccc}
\thickhline
Desig. & RA (J2000) & Dec (J2000) & Pos. err. & Probability & Magnitude & Abs. mag. & Period\\
\thickhline
\multicolumn{7}{c}{\bf SLX 1735-269} \\
\hline
 src A & 17:38:17.11 & -26:59:39.1 & 0.2\tnote{a} & 0.07 & $R = 21.31 \pm 0.12$; $I = 20.10 \pm 0.07$ & $M_I \gtrsim 3.4$ & $\lesssim 2.6$\tnote{d}\\
\hline
\multicolumn{7}{c}{\bf 3A 1742-294} \\ 
\hline
 & 17:46:05.20 & -29:30:53.3 & 1.3\tnote{c} & 0.55 & $J > 19.9$; $H > 18.6$; $K > 17.9$ & $M_K \gtrsim 2.0$ & $\lesssim 28$\tnote{e}\\
\hline
\multicolumn{7}{c}{\bf SLX 1744-299} \\ 
\hline
 src A & 17:47:25.89 & -30:00:01.6 & 0.4\tnote{a} & 0.17 & $I = 23.37 \pm 0.28; J > 19.9; H > 18.7; K > 18.0$ & $M_I \gtrsim 2.2$ & $\lesssim 24$\tnote{f}\\
\hline
\multicolumn{7}{c}{\bf SLX 1744-300} \\ 
\hline
& 17:47:26.01 & -30:02:41.8 & 0.7\tnote{b} & 0.17 & $I > 23.4; J > 19.8$; $H > 18.8$; $K > 18.0$ & $M_K \gtrsim 1.9$ & $\lesssim 65$\tnote{f}\\
%source A & 17:47:25.97 & -30:02:41.1 & 0.2\tnote{a} & 1.40 (K) & $I > 23.4; J > 20.0; H = 17.90 \pm 0.14; K = 16.75 \pm 0.14$ \\
%source B & 17:47:25.89 & -30:02:40.3 & 0.3\tnote{a} & 1.40 (K) & $I > 23.4; J > 20.0; H = 18.6 \pm 0.5; K = 18.0 \pm 0.5$ \\
\hline
\multicolumn{7}{c}{\bf GX 3+1} \\
\hline
src A & 17:47:56.06 & -26:33:48.9 & 0.2\tnote{a} & 0.55 & $J = 16.43 \pm 0.11$; $H = 15.42 \pm 0.09$; $K = 14.87 \pm 0.13$ & $M_K \gtrsim 0.2$ & $\lesssim 14$\tnote{g}\\
src B & 17:47:56.21 & -26:33:49.3 & 0.3\tnote{a} & 0.55 & $J = 19.8 \pm 0.4$; $H = 18.4 \pm 0.3$; $K = 17.9 \pm 0.3$ & \\
\hline
\multicolumn{7}{c}{\bf IGR J17505-2644} \\
\hline
 src A & 17:50:39.48 & -26:44:36.3 & 0.3\tnote{a} & 0.13 & $J > 20.3$; $H > 19.3$; $K = 18.5 \pm 0.4$ & $M_K \gtrsim 2.2$ & $\lesssim 390$\tnote{h}\\
\thickhline
\end{tabular}
%footnotesize
\begin{tablenotes}
\item [a] From optical/NIR observations, this study
\item [b] From X-ray observations, this study
\item [c] From \citet{wijnands06}
\item [d] Using $L_X$ (0.5-100 keV) from \citet{molkov05} transformed to 2-10 keV by means of relation from \citet{revnivtsev08}
\item [e] Using $L_X$ (2-10 keV) from \citet{wijnands06}
\item [f] Using $L_X$ (2-10 keV) from \citet{sidoli99}
\item [g] Assuming $L_X$ from \citet{galloway08} all comes from 2-10 keV \citep{revnivtsev08}
\item [h] Using $L_X$ (17-60 keV) from \citet{revnivtsev08} transformed to 2-10 keV by means of relation from same paper
\end{tablenotes}

\end{threeparttable}
\end{table*}

\section*{Acknowledgments} 
This research is based on observations made with ESO 3.6-m NTT telescope at
the La Silla under programmes ID 67.D-0116(A) and 079.D-0385(C), and has
made use of the VizieR catalogue access tool, CDS, Strasbourg, France. 
Authors were supported by the Russian Foundation for Basic Research, grant
10-02-00492. IZ was also supported by the RFBR grant 09-02-00032.

\bibliographystyle{mn2e}
\bibliography{iz_bib}

\begin{thebibliography}{52}
\expandafter\ifx\csname natexlab\endcsname\relax\def\natexlab#1{#1}\fi

\bibitem[{{Alexander} {et~al.}(2003){Alexander}, {Bauer}, {Brandt},
  {Schneider}, {Hornschemeier}, {Vignali}, {Barger}, {Broos}, {Cowie},
  {Garmire}, {Townsley}, {Bautz}, {Chartas}, \& {Sargent}}]{alexander03}
{Alexander}, D.~M., {et~al.} 2003, \aj, 126, 539

\bibitem[{{Bandyopadhyay} {et~al.}(1999){Bandyopadhyay}, {Shahbaz}, {Charles},
  \& {Naylor}}]{bandyopadhyay99}
{Bandyopadhyay}, R.~M., {Shahbaz}, T., {Charles}, P.~A., \& {Naylor}, T. 1999,
  \mnras, 306, 417

\bibitem[{{Belczynski} {et~al.}(2002){Belczynski}, {Kalogera}, \&
  {Bulik}}]{belczynski02}
{Belczynski}, K., {Kalogera}, V., \& {Bulik}, T. 2002, \apj, 572, 407

\bibitem[{{Bertin}(2006)}]{bertin06}
{Bertin}, E. 2006, in Astronomical Society of the Pacific Conference Series,
  Vol. 351, Astronomical Data Analysis Software and Systems XV, ed.
  C.~{Gabriel}, C.~{Arviset}, D.~{Ponz}, \& S.~{Enrique}, 112--+

\bibitem[{{Burderi} {et~al.}(2009){Burderi}, {Riggio}, {di Salvo}, {Papitto},
  {Menna}, {D'A{\`i}}, \& {Iaria}}]{burderi09}
{Burderi}, L., {Riggio}, A., {di Salvo}, T., {Papitto}, A., {Menna}, M.~T.,
  {D'A{\`i}}, A., \& {Iaria}, R. 2009, \aap, 496, L17

\bibitem[{{Cardelli} {et~al.}(1989){Cardelli}, {Clayton}, \&
  {Mathis}}]{cardelli89}
{Cardelli}, J.~A., {Clayton}, G.~C., \& {Mathis}, J.~S. 1989, \apj, 345, 245

\bibitem[{{Chenevez} {et~al.}(2006){Chenevez}, {Falanga}, {Brandt},
  {Farinelli}, {Frontera}, {Goldwurm}, {in't Zand}, {Kuulkers}, \&
  {Lund}}]{chenevez06}
{Chenevez}, J., {et~al.} 2006, \aap, 449, L5

\bibitem[{{Deacon} {et~al.}(2009){Deacon}, {Hambly}, {King}, \&
  {McCaughrean}}]{deacon09}
{Deacon}, N.~R., {Hambly}, N.~C., {King}, R.~R., \& {McCaughrean}, M.~J. 2009,
  \mnras, 394, 857

\bibitem[{{den Hartog} {et~al.}(2003){den Hartog}, {in't Zand}, {Kuulkers},
  {Cornelisse}, {Heise}, {Bazzano}, {Cocchi}, {Natalucci}, \&
  {Ubertini}}]{hartog03}
{den Hartog}, P.~R., {et~al.} 2003, \aap, 400, 633

\bibitem[{{Fabbiano}(1989)}]{fabbiano89}
{Fabbiano}, G. 1989, \araa, 27, 87

\bibitem[{{Fabbiano} {et~al.}(2007){Fabbiano}, {Brassington}, {Zezas}, {Zepf},
  {Angelini}, {Davies}, {Gallagher}, {Kalogera}, {Kim}, {King}, {Kundu},
  {Pellegrini}, \& {Trinchieri}}]{fabbiano07}
{Fabbiano}, G., {et~al.} 2007, ArXiv e-prints

\bibitem[{{Fienga} \& {Andrei}(2004)}]{fienga04}
{Fienga}, A. \& {Andrei}, A.~H. 2004, \aap, 420, 1163

\bibitem[{{Galloway} {et~al.}(2008){Galloway}, {Muno}, {Hartman}, {Psaltis}, \&
  {Chakrabarty}}]{galloway08}
{Galloway}, D.~K., {Muno}, M.~P., {Hartman}, J.~M., {Psaltis}, D., \&
  {Chakrabarty}, D. 2008, \apjs, 179, 360

\bibitem[{{Giacconi} {et~al.}(1962){Giacconi}, {Gursky}, {Paolini}, \&
  {Rossi}}]{giacconi62}
{Giacconi}, R., {Gursky}, H., {Paolini}, F.~R., \& {Rossi}, B.~B. 1962,
  Physical Review Letters, 9, 439

\bibitem[{{Gilfanov}(2004)}]{gilfanov04}
{Gilfanov}, M. 2004, \mnras, 349, 146

\bibitem[{{Gilfanov} \& {Arefiev}(2005)}]{gilfanov05}
{Gilfanov}, M. \& {Arefiev}, V. 2005, ArXiv Astrophysics e-prints

\bibitem[{{Hertz} \& {Grindlay}(1984)}]{hertz84}
{Hertz}, P. \& {Grindlay}, J.~E. 1984, \apj, 282, 118

\bibitem[{{Homan} {et~al.}(2005){Homan}, {Buxton}, {Markoff}, {Bailyn},
  {Nespoli}, \& {Belloni}}]{homan05}
{Homan}, J., {Buxton}, M., {Markoff}, S., {Bailyn}, C.~D., {Nespoli}, E., \&
  {Belloni}, T. 2005, \apj, 624, 295

\bibitem[{{Howell} {et~al.}(2001){Howell}, {Nelson}, \& {Rappaport}}]{howell01}
{Howell}, S.~B., {Nelson}, L.~A., \& {Rappaport}, S. 2001, \apj, 550, 897

\bibitem[{{Jernigan} {et~al.}(1978){Jernigan}, {Bradt}, {Doxsey}, {Dower},
  {McClintock}, \& {Apparao}}]{jernigan78}
{Jernigan}, J.~G., {Bradt}, H.~V., {Doxsey}, R.~E., {Dower}, R.~G.,
  {McClintock}, J.~E., \& {Apparao}, K.~M.~V. 1978, \nat, 272, 701

\bibitem[{{Jonker} {et~al.}(2000){Jonker}, {Fender}, {Hambly}, \& {van der
  Klis}}]{jonker00}
{Jonker}, P.~G., {Fender}, R.~P., {Hambly}, N.~C., \& {van der Klis}, M. 2000,
  \mnras, 315, L57

\bibitem[{{Kawai} {et~al.}(1988){Kawai}, {Fenimore}, {Middleditch}, {Cruddace},
  {Fritz}, {Snyder}, \& {Ulmer}}]{kawai88}
{Kawai}, N., {Fenimore}, E.~E., {Middleditch}, J., {Cruddace}, R.~G., {Fritz},
  G.~G., {Snyder}, W.~A., \& {Ulmer}, M.~P. 1988, \apj, 330, 130

\bibitem[{{Kim} \& {Fabbiano}(2004)}]{kim04}
{Kim}, D. \& {Fabbiano}, G. 2004, \apj, 611, 846

\bibitem[{{Kolb}(1993)}]{kolb93}
{Kolb}, U. 1993, \aap, 271, 149

\bibitem[{{Krivonos} {et~al.}(2007){Krivonos}, {Revnivtsev}, {Lutovinov},
  {Sazonov}, {Churazov}, \& {Sunyaev}}]{krivonos07}
{Krivonos}, R., {Revnivtsev}, M., {Lutovinov}, A., {Sazonov}, S., {Churazov},
  E., \& {Sunyaev}, R. 2007, \aap, 475, 775

\bibitem[{{Kuulkers} {et~al.}(2003){Kuulkers}, {den Hartog}, {in't Zand},
  {Verbunt}, {Harris}, \& {Cocchi}}]{kuulkers03}
{Kuulkers}, E., {den Hartog}, P.~R., {in't Zand}, J.~J.~M., {Verbunt},
  F.~W.~M., {Harris}, W.~E., \& {Cocchi}, M. 2003, \aap, 399, 663

\bibitem[{{Kuulkers} {et~al.}(2002){Kuulkers}, {Homan}, {van der Klis},
  {Lewin}, \& {M{\'e}ndez}}]{kuulkers02}
{Kuulkers}, E., {Homan}, J., {van der Klis}, M., {Lewin}, W.~H.~G., \&
  {M{\'e}ndez}, M. 2002, \aap, 382, 947

\bibitem[{{Kuulkers} \& {van der Klis}(2000)}]{kuulkers00}
{Kuulkers}, E. \& {van der Klis}, M. 2000, \aap, 356, L45

\bibitem[{{Lawrence} {et~al.}(2007){Lawrence}, {Warren}, {Almaini}, {Edge},
  {Hambly}, {Jameson}, {Lucas}, \& {Casali}}]{lawrence07}
{Lawrence}, A., {Warren}, S.~J., {Almaini}, O., {Edge}, A.~C., {Hambly}, N.~C.,
  {Jameson}, R.~F., {Lucas}, P., \& {Casali}, M. 2007, \mnras, 379, 1599

\bibitem[{{Maccarone} {et~al.}(2010){Maccarone}, {Long}, {Knigge}, {Dieball},
  \& {Zurek}}]{maccarone10}
{Maccarone}, T.~J., {Long}, K.~S., {Knigge}, C., {Dieball}, A., \& {Zurek},
  D.~R. 2010, \mnras, 834

\bibitem[{{Marshall} {et~al.}(2006){Marshall}, {Robin}, {Reyl{\'e}},
  {Schultheis}, \& {Picaud}}]{marshall06}
{Marshall}, D.~J., {Robin}, A.~C., {Reyl{\'e}}, C., {Schultheis}, M., \&
  {Picaud}, S. 2006, \aap, 453, 635

\bibitem[{{Migliari} {et~al.}(2010){Migliari}, {Tomsick}, {Miller-Jones},
  {Heinz}, {Hynes}, {Fender}, {Gallo}, {Jonker}, \& {Maccarone}}]{migliari10}
{Migliari}, S., {et~al.} 2010, \apj, 710, 117

\bibitem[{{Molkov} {et~al.}(2005){Molkov}, {Revnivtsev}, {Lutovinov}, \&
  {Sunyaev}}]{molkov05}
{Molkov}, S., {Revnivtsev}, M., {Lutovinov}, A., \& {Sunyaev}, R. 2005, \aap,
  434, 1069

\bibitem[{{Naylor} {et~al.}(1991){Naylor}, {Charles}, \& {Longmore}}]{naylor91}
{Naylor}, T., {Charles}, P.~A., \& {Longmore}, A.~J. 1991, \mnras, 252, 203

\bibitem[{{Nelemans}(2009)}]{nelemans09}
{Nelemans}, G. 2009, Classical and Quantum Gravity, 26, 094030

\bibitem[{{Priedhorsky} \& {Holt}(1987)}]{priedhorsky87}
{Priedhorsky}, W.~C. \& {Holt}, S.~S. 1987, Space Science Reviews, 45, 291

\bibitem[{{Primini} {et~al.}(1993){Primini}, {Forman}, \& {Jones}}]{primini93}
{Primini}, F.~A., {Forman}, W., \& {Jones}, C. 1993, \apj, 410, 615

\bibitem[{{Rappaport} {et~al.}(1983){Rappaport}, {Verbunt}, \&
  {Joss}}]{rappaport83}
{Rappaport}, S., {Verbunt}, F., \& {Joss}, P.~C. 1983, \apj, 275, 713

\bibitem[{{Revnivtsev} {et~al.}(2008){Revnivtsev}, {Lutovinov}, {Churazov},
  {Sazonov}, {Gilfanov}, {Grebenev}, \& {Sunyaev}}]{revnivtsev08}
{Revnivtsev}, M., {Lutovinov}, A., {Churazov}, E., {Sazonov}, S., {Gilfanov},
  M., {Grebenev}, S., \& {Sunyaev}, R. 2008, \aap, 491, 209

\bibitem[{{Russell} {et~al.}(2007){Russell}, {Fender}, \& {Jonker}}]{russel07}
{Russell}, D.~M., {Fender}, R.~P., \& {Jonker}, P.~G. 2007, \mnras, 379, 1108

\bibitem[{{Sarazin} {et~al.}(2003){Sarazin}, {Kundu}, {Irwin}, {Sivakoff},
  {Blanton}, \& {Randall}}]{sarazin03}
{Sarazin}, C.~L., {Kundu}, A., {Irwin}, J.~A., {Sivakoff}, G.~R., {Blanton},
  E.~L., \& {Randall}, S.~W. 2003, \apj, 595, 743

\bibitem[{{Sidoli} {et~al.}(1999){Sidoli}, {Mereghetti}, {Israel},
  {Chiappetti}, {Treves}, \& {Orlandini}}]{sidoli99}
{Sidoli}, L., {Mereghetti}, S., {Israel}, G.~L., {Chiappetti}, L., {Treves},
  A., \& {Orlandini}, M. 1999, \apj, 525, 215

\bibitem[{{Siuniaev} {et~al.}(1991){Siuniaev}, {Churazov}, {Gil'Fanov},
  {Pavlinskii}, {Grebenev}, {Dekhanov}, {Kuznetsov}, {Iyamburenko}, {Skinner},
  \& {Patterson}}]{sunyaev91}
{Siuniaev}, R.~A., {et~al.} 1991, Advances in Space Research, 11, 177

\bibitem[{{Skinner} {et~al.}(1990){Skinner}, {Foster}, {Willmore}, \&
  {Eyles}}]{skinner90}
{Skinner}, G.~K., {Foster}, A.~J., {Willmore}, A.~P., \& {Eyles}, C.~J. 1990,
  \mnras, 243, 72

\bibitem[{{Skinner} {et~al.}(1987){Skinner}, {Willmore}, {Eyles}, {Bertram}, \&
  {Church}}]{skinner87}
{Skinner}, G.~K., {Willmore}, A.~P., {Eyles}, C.~J., {Bertram}, D., \&
  {Church}, M.~J. 1987, \nat, 330, 544

\bibitem[{{Skrutskie} {et~al.}(2006){Skrutskie}, {Cutri}, {Stiening},
  {Weinberg}, {Schneider}, {Carpenter}, {Beichman}, {Capps}, {Chester},
  {Elias}, {Huchra}, {Liebert}, {Lonsdale}, {Monet}, {Price}, {Seitzer},
  {Jarrett}, {Kirkpatrick}, {Gizis}, {Howard}, {Evans}, {Fowler}, {Fullmer},
  {Hurt}, {Light}, {Kopan}, {Marsh}, {McCallon}, {Tam}, {Van Dyk}, \&
  {Wheelock}}]{skrutskie06}
{Skrutskie}, M.~F., {et~al.} 2006, \aj, 131, 1163

\bibitem[{{van Paradijs} \& {McClintock}(1994)}]{paradijs94}
{van Paradijs}, J. \& {McClintock}, J.~E. 1994, \aap, 290, 133

\bibitem[{{Wachter} {et~al.}(2005){Wachter}, {Wellhouse}, {Patel}, {Smale},
  {Alves}, \& {Bouchet}}]{wachter05b}
{Wachter}, S., {Wellhouse}, J.~W., {Patel}, S.~K., {Smale}, A.~P., {Alves},
  J.~F., \& {Bouchet}, P. 2005, \apj, 621, 393

\bibitem[{{Wijnands} {et~al.}(2006){Wijnands}, {in't Zand}, {Rupen},
  {Maccarone}, {Homan}, {Cornelisse}, {Fender}, {Grindlay}, {van der Klis},
  {Kuulkers}, {Markwardt}, {Miller-Jones}, \& {Wang}}]{wijnands06}
{Wijnands}, R., {et~al.} 2006, \aap, 449, 1117

\bibitem[{{Wilson} {et~al.}(2003){Wilson}, {Patel}, {Kouveliotou}, {Jonker},
  {van der Klis}, {Lewin}, {Belloni}, \& {M{\'e}ndez}}]{wilson03}
{Wilson}, C.~A., {Patel}, S.~K., {Kouveliotou}, C., {Jonker}, P.~G., {van der
  Klis}, M., {Lewin}, W.~H.~G., {Belloni}, T., \& {M{\'e}ndez}, M. 2003, \apj,
  596, 1220

\bibitem[{{Zolotukhin}(2009)}]{zolotukhin_atel_09}
{Zolotukhin}, I. 2009, The Astronomer's Telegram, 2032, 1

\bibitem[{{Zolotukhin} {et~al.}(2010){Zolotukhin}, {Revnivtsev}, \&
  {Shakura}}]{zolotukhin10a}
{Zolotukhin}, I.~Y., {Revnivtsev}, M.~G., \& {Shakura}, N.~I. 2010, \mnras,
  401, L1

\end{thebibliography}

\label{lastpage}

\end{document}